\documentclass{WileyMSP-template}

\usepackage{graphicx}
\usepackage{epstopdf}
\usepackage{ragged2e}
\usepackage{amsmath,amsfonts,amssymb,amsthm}
\usepackage{cite}
\usepackage{float}
\usepackage{lineno}
\usepackage{xcolor}

\begin{document}

\pagestyle{fancy}
\rhead{\includegraphics[width=2.5cm]{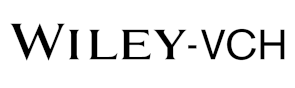}}

\title{Building Intelligence in the Mechanical Domain -- Harvesting the Reservoir Computing Power in Origami to Achieve Information Perception Tasks}
\maketitle

% Author: Please give full first and last names for authors and include * after the name of all corresponding authors
\author{Jun Wang* and Suyi Li}

% Affiliations: Please provide academic titles (Prof. or Dr.) for all authors where applicable, and include an institutional email address for all corresponding authors
\begin{affiliations}

Department of Mechanical Engineering, Virginia Tech\\
181 Durham Hall, 1145 Perry Street, Blacksburg, VA 24061, USA\\[0.1in]

*Correspondent email address: \texttt{junw@vt.edu}

\end{affiliations}

% Keywords: Please provide a minimum of three and a maximum of seven keywords, separated by commas

\begin{abstract}
%\justifying
In this paper, we experimentally examine the cognitive capability of a simple, paper-based Miura-ori --- using the physical reservoir computing framework --- to achieve different information perception tasks. The body dynamics of Miura-ori (aka. its vertices displacements), which is excited by a simple harmonic base excitation, can be exploited as the reservoir computing resource. By recording these dynamics with a high-resolution camera and image processing program and then using linear regression for training, we show that the origami reservoir has sufficient computing capacity to estimate the weight and position of a payload. It can also recognize the input frequency and magnitude patterns. Furthermore, multitasking is achievable by simultaneously applying two targeted functions to the same reservoir state matrix. Therefore, we demonstrate that Miura-ori can assess the dynamic interactions between its body and ambient environment to extract meaningful information --- an intelligent behavior in the mechanical domain. Given that Miura-ori has been widely used to construct deployable structures, lightweight materials, and compliant robots, enabling such information perception tasks can add a new dimension to the functionality of such a versatile structure.
\end{abstract}

\section{Introduction}

%\linenumbers
%\justifying

We are witnessing an ever-increasing demand for the next generation of multi-functional structures and material systems that can behave intelligently according to their mission needs and dynamic working conditions. 
Ideally, these intelligent systems should observe their environment, extract critical information from sensory inputs, learn from past experiences, decide on the action plan, and execute control commands --- in a highly integrated and distributed setup and in real time.
Such intelligence is traditionally implemented in the digital domain with the help of, for example, onboard computers. However, there has been an increasing interest in offloading and distributing some of the intelligence into the physical domain without a centralized digital computer or even without any electronics \cite{sitti2021physical,iida2022timescales,pfeifer2004embodied}. 
In this regard, animals have offered us many inspirations  \cite{mengaldo2022concise, laschi2016lessons} as they can outsource many information processing and locomotion control tasks from their brain to the body --- by exploiting its physical morphology, mechanics, and neuro-muscular structure.
For example, the octopus' hydrostat-muscular arm has an elaborated muscular layout with a distributed sensory-neural network so that it can accomplish many complex locomotions and manipulations without the direct involvement of its brain \cite{nakajima2013soft}. 
Inspired by these lessons from nature, researchers developed many new strategies for achieving (or mimicking) intelligent behaviors in the mechanical domain. For example, one can use the multi-stability embedded in soft materials to sequence locomotion gaits \cite{bhovad2019peristaltic}, replace digital circuitry with fluidic ones for logic operation in entirely soft substrates \cite{hevia2022towards}, and exploit wave phenomena to achieve computation directly in metamaterial systems \cite{silva2014performing} .  
Overall, achieving intelligence using mechanical components (we refer to as ``mechano-intelligence'' hereafter \cite{mcevoy2015materials}) could bring significant advantages such as lower power consumption, less analog-digital conversion, higher overall speed, and better survivability in harsh environments. 
%
% in the physical domain and "outsourcing" parts of the intelligent tasks directly to the mechanical components like integrated devices \cite{fang2016pattern, farrow2016morphological}, responsive materials \cite{fang2017designing, asadnia2016biological, meng2020mechanicaly}, or nonlinear mechanics \cite{bhovad2021physical, gorissen2019hardware, berwind2018hierarchical, vasios2020harnessing}. 

\medskip
Nonetheless, intelligence is a broad concept consisting of many interconnected facets, including (and not limited to) information perception, decision-making, learning with memory, and command execution. The studies mentioned above are limited to a particular intelligence task and mostly focused on command execution. We still need a versatile foundation for achieving multiple intelligent tasks of different natures. 

\begin{figure}[h]
    \centering
    \includegraphics[scale=0.95]{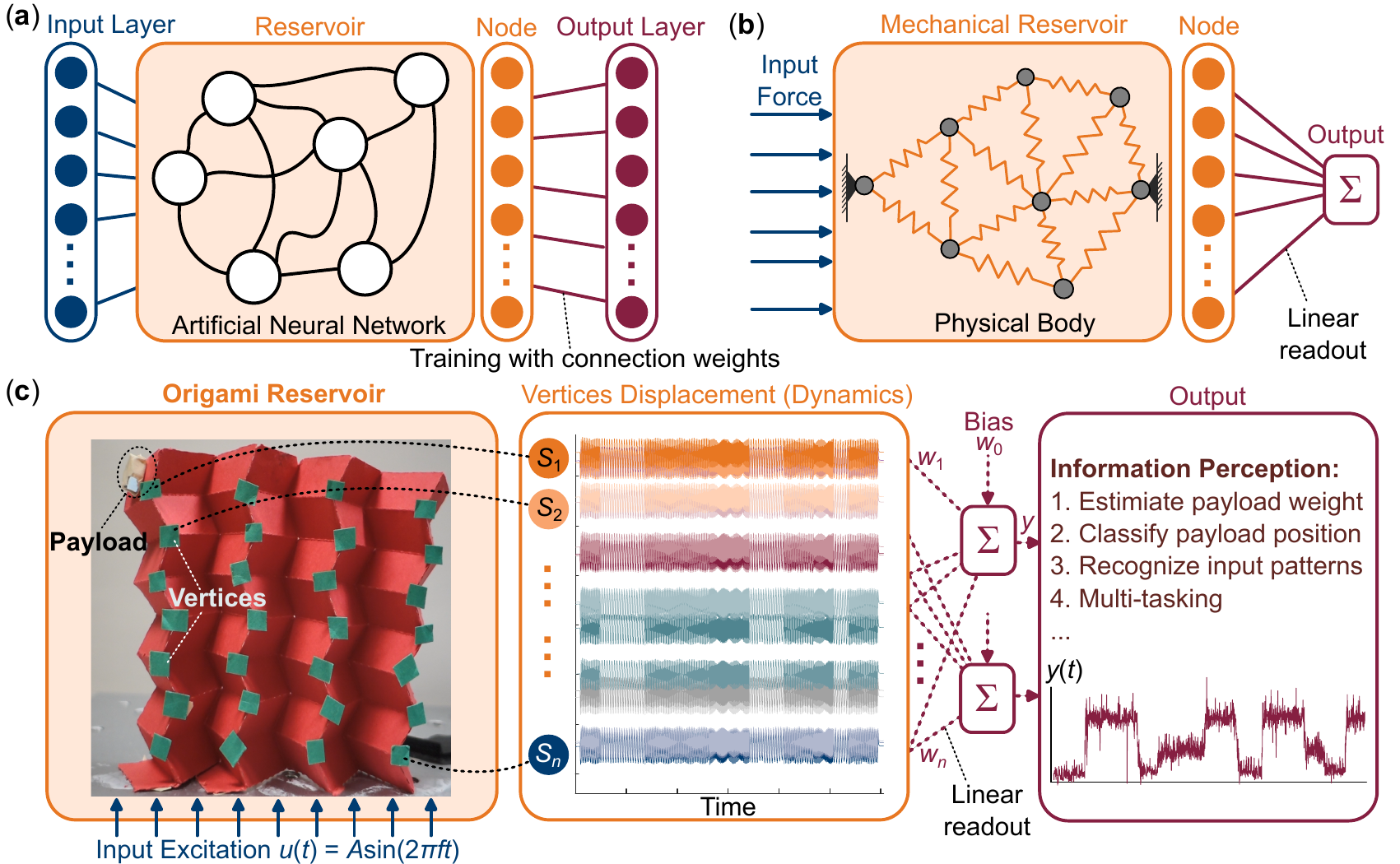}
    \caption{The overall concept of physical reservoir computing and its implementation in this study: (\textbf{a}) The conceptual setup of a generic reservoir computer, where only the connection weights between the reservoir's nodal response and the output layer are trained. (\textbf{b}) A well-studied example of a physical reservoir computer that uses a random network of mass and nonlinear spring as the computing kernel \cite{hauser2011towards}. (\textbf{c}) Overall setup for the origami-based reservoir in this study. Here, a very simple, paper-folded origami is the physical kernel, the dynamic vibrations of its vertices are the reservoir states, and the target intelligent behaviors are information perception tasks. In detail, the folded Miura-ori is attached to the shaker receiving vibration signal. Then displacements for 28 vertices (attached with green cards) are recorded, acting as reservoir states. A group of linear readout weights could be obtained by training this set of vectors with linear regression and setting the target output as perception information (e.g. payload weight, payload position, input patterns). This set of readout weight then could be used to predict different information received by Miura-ori.  }
    \label{fig:intro}
\end{figure}

\medskip
To this end, this study examines the potential of {\it physical reservoir computing} (PRC) as such a foundation. 
Reservoir computing is a branch within the discipline of recurrent artificial neural networks. In a reservoir computer, the interconnection weights inside the neural network's kernel remain fixed, and only output weights are trained to reach the targeted output (\textbf{Figure \ref{fig:intro}}a) \cite{schrauwen2007overview}. Such a fixed neural net kernel is called the ``reservoir.''  Reservoir computing has been widely used in time-series prediction tasks like robotic motion planning \cite{wyffels2009design} and text recognition \cite{abreu2020role}. More importantly, since the reservoir does not change during training, one can use a physical body as the reservoir and harvest its nonlinear and high-dimensional dynamic responses as the computational resource \cite{hauser2011towards} (essentially, the physical body itself becomes the neural network).
For example, a recurrently connected mass-spring-damper system could act as a physical reservoir and achieve different machine learning tasks like emulation, pattern generation, and even text recognition (Figure 1b). Following up on this idea, researchers used many other physical systems to achieve reservoir computing, including soft robots \cite{li2012behavior, nakajima2017muscular, horii2021physical}, tensegrity structure \cite{caluwaerts2013locomotion, fujita2018environmental}, and pneumatically driven apparatus \cite{kawase2021pneumatic}.

\medskip
A key advantage of physical reservoir computing is its simplicity and flexibility to achieve different tasks --- so it can become the foundation for multi-faceted mechano-intelligence. In the physical reservoir, only a group of linear readout weights are needed for its state vectors (e.g., spring lengths in Figure \ref{fig:intro}b) to achieve the targeted output, and one can train these readout weights with simple linear regressions. This setup means the nonlinear dynamics of the physical body are most responsible for complex computation.
Moreover, one can apply separate sets of readout weights to the same reservoir to achieve different machine-learning tasks concurrently.
Finally, PRC is also a powerful method for edge computing devices \cite{yamane2018dimensionality} since it can store memory and process information simultaneously without time delay.

\medskip
Therefore, this study examines the use of physical reservoir computing as the foundation to achieve multiple and complex intelligent tasks. More specifically, we use a very simple, paper-folded Miura-ori as the physical kernel (Figure \ref{fig:intro}c) and show that it has sufficient reservoir computing capacity by projecting input signals (e.g., forces from the environment) to a high-dimensional state space (aka. nonlinear vibrations in its body).  As a result, the Miura-ori can analyze the dynamic interactions between its body and the surrounding environment to extract valuable knowledge (aka. information perception) -- a vital but sparsely studied component of intelligence in the mechanical domain.
%55Origami is an ideal candidate for acting as the kernel to accomplish physical reservoir computing and, eventually, mechano-intelligent tasks. It satisfies three essential requirements for a mechanical system to perform as a reservoir \cite{bhovad2021physical}. First, origami is fundamentally high-dimensional because of its distributed compliance in the facets and creases. Secondly, origami is inherently nonlinear due to the large-amplitude deformations via folding. Finally, origami has the ``separation property,'' which arises from the deterministic nature of structural dynamics. 
%
%
% In our previous study, we have shown via numerical simulation that origami has sufficient computation power to emulate high-order nonlinear dynamical systems and generate stable limit cycles for crawling locomotion gait.

\medskip
We use extensive experimentation to show that an origami reservoir can ``observe'' and ``analyze'' its mechanical vibration responses and achieve three information perception tasks.
1) The origami reservoir can accurately estimate the weight of payloads applied at its top edge, 2) it can classify the position of these payloads, and 3) it can recognize different input signal patterns (frequency and amplitude). 
Moreover, we experimentally show that the origami reservoir can simultaneously achieve any combination of the two tasks (aka. multi-tasking). 
One can achieve these tasks simply by extracting origami's displacement fields from videos and training the corresponding readout weights with linear regressions. 

% The key advantage of this approach is viewing high nonlinear and high-dimensional response as a solution rather than the source of the problem. Soft properties, which are typically unwanted in robotic control, are utilized as potential computational resources for computing \cite{nakajima2018exploiting}, memory \cite{inubushi2017reservoir}, and control \cite{li2012behavior}. 
%
% Among these mechano-intelligence strategies, the nonlinear and high-dimensional structural dynamic responses have shown surprisingly versatile computing capabilities. 
% For example, carefully-designed metamaterials and devices can conduct mathematical operations like differentiation \cite{silva2014performing, zuo2018acoustic} and Boolean logic \cite{song2019additively, waheed2020boolean} via wave dynamics or multi-stability. 

\medskip
In what follows, we first explain the experimental setup, including the physical platform and computing framework for origami reservoir computing. Next, we detail the results regarding the information perception tasks. We further study the influence of reduced computing dimension on intelligence performance. Finally, we end this paper with a summary and discussion.

\section{Physical Reservoir Computing Setup}
\subsection{Physical Kernel}
\begin{figure}[h]
    \centering
    \includegraphics[scale=0.95]{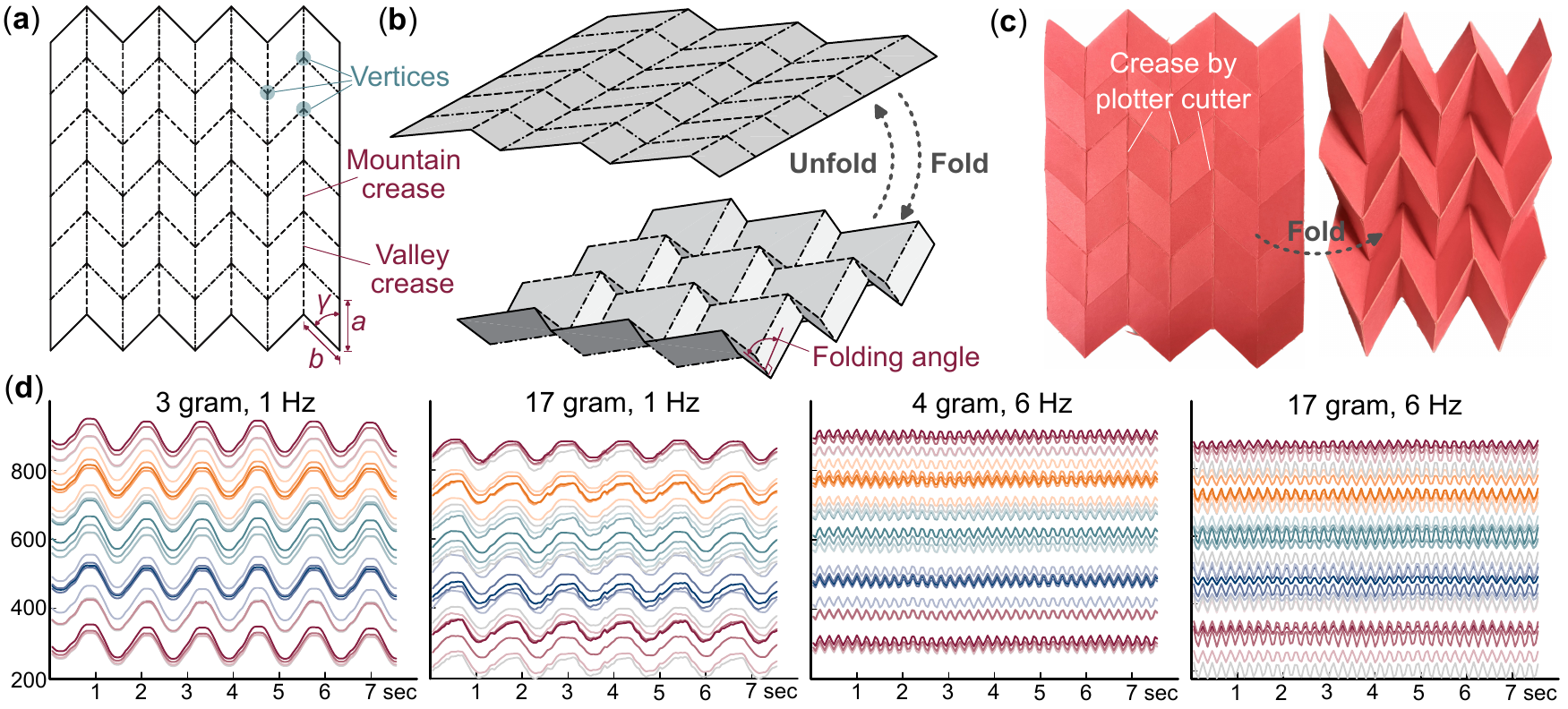}
    \caption{Setting up the origami reservoir computing. (a) The crease geometry of a Miura-ori shows the tessellation of mountain folds (dashed line) and valley folds (dash-dotted line). (b) 3D illustration of the folding process. Note that the folding angle is the dihedral angle between two quadrilateral facets. (c) Paper-based Miura-ori on plotter cutter and after folding. (d) Vertices (or nodal) displacements recorded by the camera. The payload is 3 and 17 grams, and the input frequency is 1 or 6 Hz. Different color lines represent the different vertices in Miura-ori. }
    \label{fig:setup}
\end{figure}

In all experiments, we use a traditionally folded Miura-ori sheet made from 1.3mm thin print paper as the computation kernel (\textbf{Figure \ref{fig:setup}}a). Miura-ori is a classical origami pattern that has been used as the backbone of many deployable structures \cite{deleo2020origami}, multi-functional materials \cite{wu2021multifunctional}, and compliant robotics \cite{rus2018design} . It is a periodic tessellation of unit cells consisting of four identical quadrilateral facets connected by mountain and valley creases. 
We first perforate the crease pattern into the print paper on a mechanical plotter machine (GraphicTech FCX4000-60) and manually fold the Miura-ori until the dihedral ``folding'' angle between facets is between 45\degree to 60\degree.  
Then we place the folded Miura-ori on a large-stroke exciter (APS 113) and fully fix origami to the exciter table on its left bottom corner. This way, the origami can receive base excitation $u(t)$ generated via a National Instrument DAQ system with a power amplifier (Figure \ref{fig:intro}c).

\medskip
During the tests, we attach small green markers to Miura-ori's vertices--points where the crease lines meet---and record their displacements using a video camera at 25 frames per second with 720p resolution. The size of the origami reservoir (or the number of green markers) is, therefore, $4\times7$.
Then, by post-processing the video footage in MATLAB, we obtain all vertical displacements of the 28 vertices (or nodes), labeled as $s_1, \ldots, s_n$ shown in Figure \ref{fig:intro}(c). These data are used as the reservoir state vectors for information perception tasks. With linear regression, readout weights ($w_0, w_1, \ldots, w_n$) can be obtained to get desired output.

\subsection{Computing Framework}
Since the origami body is the fixed reservoir kernel, training is reduced to finding a set of static linear readout weight $W_{out}=[w_0, w_1, \ldots, w_n]^\intercal$ via simple linear regression:
\begin{equation}
    \label{eq:LR algorithm}
      {{W}_{out}}={{[\mathbf{I} \  \mathbf{S}]}^{+}}\widehat{y}={{\overline{\Phi}}^{+}}\widehat{y},
\end{equation}
where matrix $\mathbf{S}$ is compiled by assembling reservoir state ${{s}_{j}}$ at every training time steps, $\mathbf{I} $ is a column of ones for calculating bias term. ${{[.]}^{+}}$ refers to the Moore–Penrose pseudo-inverse to accommodate non-square matrices. $\widehat{y}$ represents reference signals in each time step, and it becomes a matrix $\widehat{Y}$ when more than one reference is provided for multi-tasking. $\widehat{y}$ is the only variable that needs to be adjusted manually in training for different tasks.

\medskip
With obtained readout weights, the reservoir output (shown in Figure \ref{fig:intro}c) is:
\begin{equation}
    \label{eq:reservoir output}
      y(t)=w_0+\sum\limits_{i=1}^N w_i s_i.
\end{equation}

\medskip
The performance is evaluated based on the rooted mean squared error (RMSE):
\begin{equation}
    \label{eq:RSME}
      RMSE=\sqrt{\frac{1}{M}\sum\limits_{j=1}^{M}{{{\left({{y}_{j}}-\overset{\wedge }{\mathop{{{y}_{j}}}}\,\right)}^{2}}}},
\end{equation}
where $M$ is the number of the video frame in evaluation, ${{y}_{j}}$ is the reservoir output in the \textit{j}\textsuperscript{th} frame, and $\overset{\wedge}{\mathop{{{y}_{j}}}}\,$ is the target value in the \textit{j}\textsuperscript{th} frame.

\subsection{Dynamic Responses}
The input signal $u(t)$ in the information perception tasks is a sinusoidal function:
\begin{equation} \label{eq:input signal}
      u(t)=A \sin (2\pi f t),
\end{equation}
where $A$ is excitation amplitude, $f$ is the input frequency, and $t$ is the elapsed time in seconds. 
We conducted multiple experiments, each with a unique combination of payload weight, payload position, and input frequency. 
The payload, made of small magnets, weighs from 3 to 18 grams (16 different levels). For reference, the origami weighs 6 grams. Its position varies from the upper left corner of the origami reservoir (labeled as ``a'') to the upper right corner (labeled as ``h''). The input frequency is set between 1 and 6 Hz (7 different levels). Therefore, 896 groups of reservoir state vectors ($16\times 8\times7$) are collected to comprehensively investigate the best training method and corresponding intelligent task performance.
In every test, vibratory input lasts for 15 seconds. The first and last 5 seconds of data (125 frames) are discarded to eliminate transient responses, and the remaining displacement data are used for reservoir training and testing.

\medskip
Figure \ref{fig:setup}(d) shows four typical time responses of the 28 vertices (aka. state vectors). In these tests, the payload is located at the top left corner of origami; the payload weights and input frequencies are $<$3g, 1Hz$>$, $<$17g, 1Hz$>$, $<$3g, 6Hz$>$, and $<$17g, 6Hz$>$. 
%
% The vertical axis represents pixel coordinates in each video frame rather than actual displacements. 
%
Comparing these data shows that the origami vertices' displacements are close to the excitation signal $u(t)$ when the input frequency is low at 1 Hz and the payload weight is small at 3 grams. In this testing condition, the Miura-ori behaves like a rigid body with minimal folding deformation. 
However, when the Miura-ori carries a much heavier payload at 17 grams (about three times its weight), its vertices displacements deviate from the input signal and show a much more significant nonlinearity. Such nonlinearity originates from the non-uniform deformation from facet bending \cite{li2019architected}, and it is required for successful reservoir computing \cite{hauser2011towards}.
It is worth noting that we test payloads up to 18 grams. Any heavier weights could generate chaotic responses and even buckles the Miura-ori.

\medskip
On the other hand, when the input frequency increases to 6Hz, the nonlinearity is not captured clearly in the recorded data even if the payload is heavy and the non-uniform dynamic deformation still exists. 
Two factors contribute to the significantly weaker nonlinearity at higher input frequencies. First, the output amplitude from the shaker naturally decreases as the frequency increases due to its power limit. Secondly, the data sampling rate --- the camera's frame per second (FPS) setting --- is 25, meaning that the camera can only capture about 4 data points per cycle with a high input frequency of 6 Hz. Such a relatively low sampling rate might lead to the loss of some critical information. 
Therefore, even though a relatively high-frequency actuation might be beneficial to achieve high computing power in the physical reservoir, we are still constrained by sensors' sensitivity and operating frequency. 

\section{Information Processing by Origami Reservoir}
\subsection{Task 1: Payload Weight Estimation}
In this section, we investigate whether the origami reservoir can directly predict the weight of its payload by assessing its influence on the time response. To train the origami to achieve this task, we need to assemble a state vector matrix by combining the reservoir state responses from two different ``training'' payloads. That is $\mathbf{S} =[\mathbf{S}^{m_1, pa}; \mathbf{S}^{m_n, pa}],$ where $\mathbf{S}^{m_1, pa}$ is 5 seconds of state vector matrix from placing the 3-gram payload on position ``a'' (upper left corner of the origami), and $\mathbf{S}^{m_n, pa}$ is another 5 seconds of state vector matrix from the other payload. The corresponding target output is a piece-wise step function whose value equals the training payload weight in that:
\begin{equation}
    \label{eq:task 1 target}
    {{\widehat{y}}_{1}}(t)=
    \begin{cases}
      {{m}_{1}=3}&(0<t<5)\\
      {{m}_{n}}&(5\le t<10)\\ 
    \end{cases}.
\end{equation}

\medskip
The readout weight ${{W}_{payload}}$ is then determined based on Eq. (1), and all other state vectors (not used for training) are used for evaluating the prediction performance. 
The final predicted payload is then the average ${{y}_{j}}$ over 125 frames of test data. 

\begin{figure}[t]
    \centering
    \includegraphics[scale=0.95]{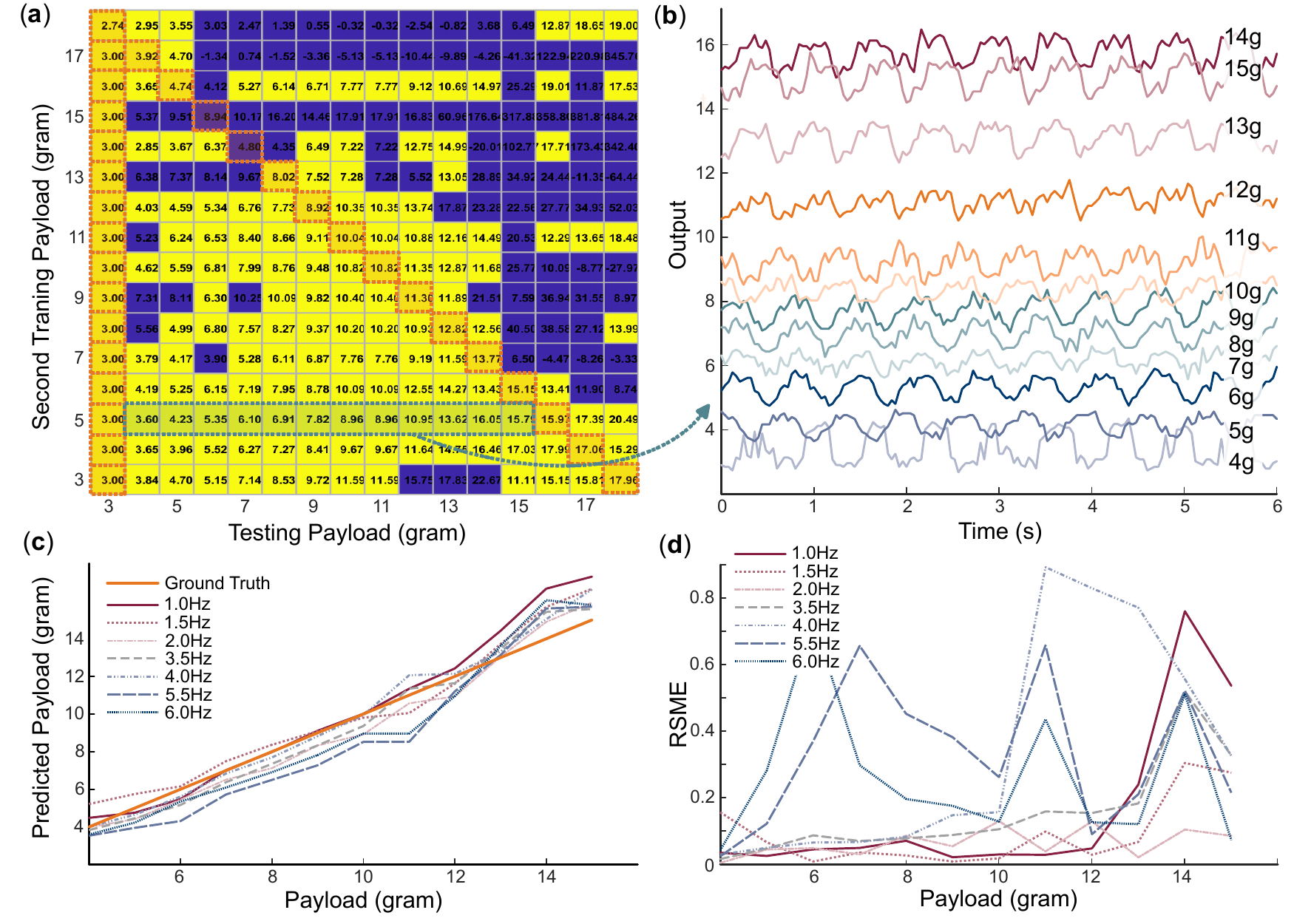}
    \caption{Estimating payload weight with different input excitation frequencies. (a) Summary of the origami's reservoirs predictions on different payload weights based on different training setups. Each row in this matrix corresponds to a unique selection of the two training masses (highlighted by orange dashed lines). For example, the second row shows the results from training with the 3 and 4-gram payloads, and the last row is from training with the 3 and 18-gram payloads. Each column in this matrix shows the origami's prediction corresponding to a different payload, so the value inside each block is the predicted payload weight, and the background color is set as yellow if the error is less than 30\%. (b) The reservoir's output corresponding to payloads from 4 to 15 grams, while readout weight is trained with the 3 and 16-gram payloads. (c) Prediction results under different excitation frequencies. The solid orange line is the ground truth. (d) The root-mean-square error of prediction with different frequencies.}
    \label{fig:weight}
\end{figure}

\medskip
Here, to rigorously assess the best training approach, we choose the second training mass $m_n$ between 4 and 18 grams, and then survey the corresponding payload weight estimation performance (the input frequency is fixed at 3 Hz). The matrix plot in \textbf{Figure \ref{fig:weight}}(a) summarizes the results. Every row in this matrix represents the results based on a combination of two training masses, highlighted by the orange dashed lines. 
For example, in the third to last row, we select the 3-gram payload as the first training mass and the 16-gram one as the second training mass to obtain the readout weights. Then, this set of readout weights is applied to the reservoir state vectors from other payloads to estimate their weight. Correspondingly, the rest of this row is the estimation result. 
Figure \ref{fig:weight}(b) shows the reservoir outputs from this 3 and 16-gram payload training. The reservoir output clearly shows separability (i.e., a small difference in the input would result in a significant change in the output so that different inputs give different outputs). This is a prerequisite for a successful reservoir.

\medskip
In the matrix plot, we mark the block with yellow if the estimation error is less than 30\%, which is considered a successful estimation, and blue otherwise. 
By carefully surveying the results from all selections of training weights, one can see that the successful cases concentrate in the lower triangle of the matrix, indicating that the origami reservoir can better predict unknown payload weights if they are within the range of two training payloads. 
The prediction becomes more reliable when the second training payload is heavier, which is likely the benefit of the more nonlinear and high-dimensional responses discussed earlier. 
In addition, an unsymmetrical setup (e.g., payload at the corner of origami) gives a better prediction than a symmetric one (e.g., payload at the center of origami). 

\medskip
The influence of input frequency on prediction accuracy is also essential to be understood. To this end, we fix the training payloads to be 3 and 16 grams at the upper left corner of origami --- the favored training setup --- and increase the input frequency from 1 to 6 Hz (1, 1.5, 2, 3.5, 4, 5.5, and 6Hz). For each tested input frequency, we obtain a separate set of readout weights from the two training payloads and apply it to predict all other payload weights (from 4 to 15 grams). Figure \ref{fig:weight}(c) summarizes all the reservoir predictions, and Figure \ref{fig:weight}(d) shows the corresponding root mean square errors (RMSE).
First, all predictions under different input frequencies show good consistency with the ground truth. 
However, some discrepancies still exist. The origami reservoir underestimates when the true payload weight is less than two times its weight, especially when the input frequency is low in the 1-3.5 Hz range.
In contrast, the reservoir overestimates the payload weight when the payload is heavier than 12 grams at higher input frequencies. This error might be caused by information loss due to the sampling rate, as we discussed before. 
Overall, the average rooted mean square error is 0.364g for all tested input frequencies.

\subsection{Task 2: Payload Position Classification}

\begin{figure}[t]
    \centering
    \includegraphics[]{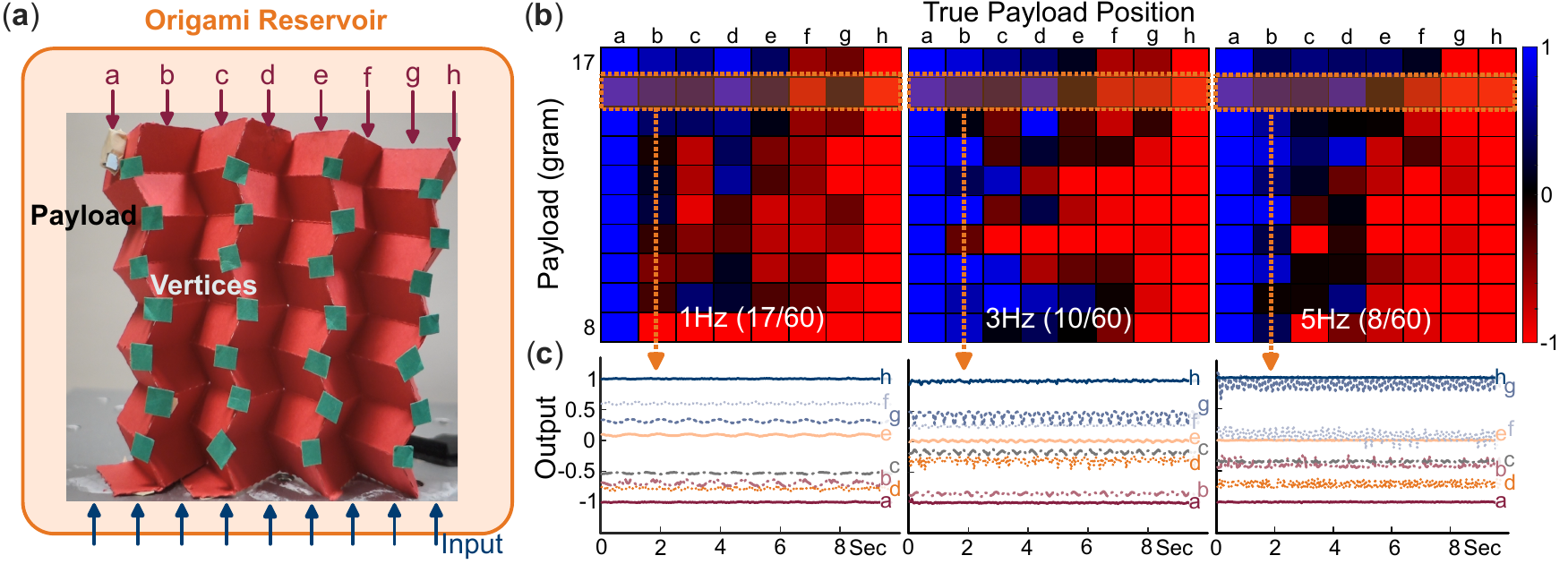}
    \caption{Results of payload position classification task under different payload and input frequencies (1Hz, 3Hz, 5Hz). (a) The definition of positions ``a'' to ``h'' on the origami body. (b) Color block diagram summarizing the origami reservoir's prediction on payload position. The payload is classified to be on the left half of origami if the output is less than 0 (blue color) and, otherwise, the right half (red color). Note that the first and last columns are training results, while columns 2 to 7 are the classification results. Each row represents the results from a different payload weighing 8 to 17 grams. (c) Example reservoir outputs when the payload is 16 grams, and the base excitation input frequencies are 1, 3, and 5 Hz, respectively.}
    \label{fig:position}
\end{figure}

For this task, we investigate whether the origami reservoir can predict the position of different payloads on its body. We use a similar setup as the payload weight estimation task to obtain the reservoir state vector matrices. For training, we assemble state vectors from the same payload but in different positions ``a'' (upper left corner) and ``h'' (upper right corner, \textbf{Figure \ref{fig:position}}a). That is $\mathbf{S} =[{\mathbf{S}^{m_i, pa}};{\mathbf{S}^{m_i, ph}}]$, and each state vector matrix component is 5 seconds long. The target output is again a piece-wise constant step function:
\begin{equation}
    \label{eq:task 2 target}
    {{\widehat{y}}_{2}}(t)=
    \begin{cases}
      -1&(0<t<5)\\
      1&(5\le t<10)\\ 
    \end{cases}.
\end{equation}

\medskip
Based on this training setup, we obtain a unique set of readout weights ${{W}_{position}}$ for each payload and then apply it to other state vectors from this payload located at positions ``b'' to ``g.'' Since the targeted output is -1 for position ``a'' and 1 for position ``g,'' the averaged reservoir outputs fall into a range of [-1, 1] when the payload is at these intermediate positions. Therefore, the payload is classified as on the origami's left side when the averaged reservoir output is less than 0 and on the right side otherwise.

\medskip
Figure \ref{fig:position}(b) summarizes the results of the position classification task from three different input frequencies and ten different payloads (ranging from 8 to 17 grams). We find that the origami reservoir could not provide reliable results when the payload's weight is roughly the same as or less than the origami (6 grams). This is likely because Miura-ori deforms relatively uniformly with small external force, so it cannot distinguish different payload positions.

\medskip
We collect all successful cases with payload positions ranging from ``b'' to ``g''  in Figure \ref{fig:position}(b). The overall error rate ---corresponding to 1, 3, and 5 Hz input frequencies --- is 17/60 (22\%), 10/60 (17\%), and 8/60 (13\%), respectively. Therefore, increasing input frequency gives higher accuracy. Figure \ref{fig:position}(c) shows example outputs from the reservoir under three frequencies, all carrying a 16-gram payload. These outputs clearly show the characteristic separation property (aka. different payload positions generate significantly different reservoir outputs) even though the predicted payload position is sometimes inaccurate. Overall, the origami reservoir can classify the payload position more accurately when these payloads are heavier (around three times the origami's weight) and located far away from the origami's center. 

\medskip
Since a relatively low input frequency is desirable for payload weight estimation but a higher frequency works better for payload position classification, an input frequency of around 2 to 3 Hz gives a balanced performance between these two tasks.

\subsection{Task 3: Input Pattern Recognition}
Input signal recognition and classification is an essential component of intelligent behavior, and it is the prerequisite for more complex learning, decision-making, and control tasks. To this end, we investigate whether the origami reservoir can learn and memorize different input frequency and amplitude patterns from the underlying shaker. If origami has such a capability, it should recognize unknown combinations of these input parameters with sets of pre-learned readout weights.

\begin{figure}[b!]
    \centering
    \includegraphics[scale=0.95]{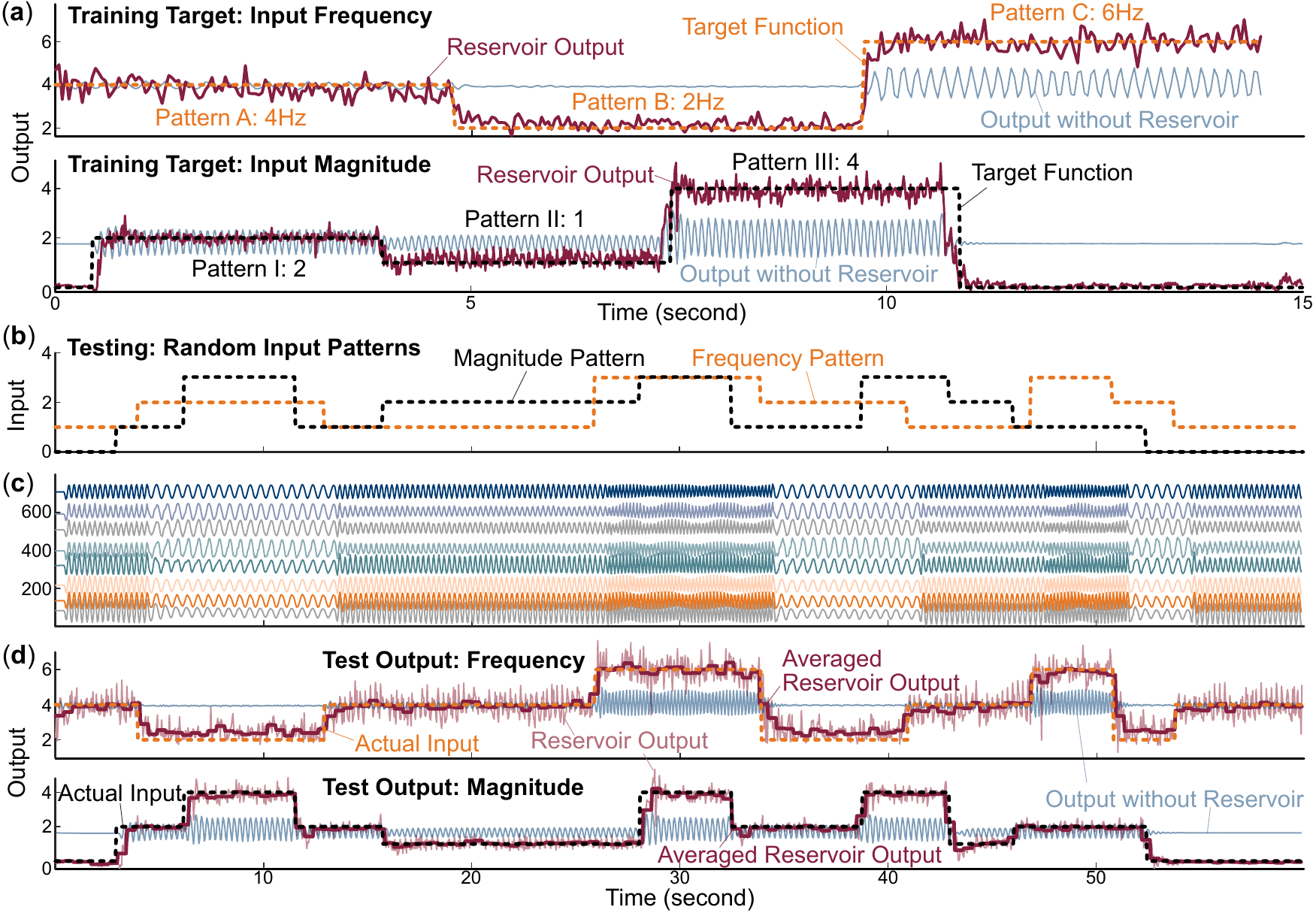}
    \caption{Identification of input patterns with different frequencies and amplitudes: (a) Training for input pattern recognition regarding frequency (up) and magnitude (below). The dashed lines are target functions for training. They are constant piece-wise functions whose value equals the input frequency magnitude (4, 2, and 6Hz) or input magnitude (level 2, 1, 4 in the shaker controller), respectively. The solid red line is the training results with the reservoir, while the blue line is the training output without involving the origami reservoir. (b) Two input patterns for testing, with random combinations of three different frequency or amplitude patterns, respectively. (c) Example data showing the displacements of 8 vertices under the testing input frequency pattern shown in (b). (d) The results of the input pattern recognition task for frequency (up) and amplitude (below), respectively. Dashed lines are the true input patterns, and solid red lines are the origami reservoir's predictions.}
    \label{fig:input}
\end{figure}

\medskip
First, we train the origami reservoir to identify input frequencies. We set up three different input frequency patterns -- 4 Hz as pattern A, 2 Hz as pattern B, and 6 Hz as pattern C -- while the amplitude is constant at ``level 2'' in the shaker controller. For the training, we place the 6-gram payload at position ``a'' to induce sufficiently nonlinear dynamic responses. Then we excite the origami by these three patterns in a sequence, each for 5 seconds, and collect the corresponding reservoir state vectors $\mathbf{S} =[{\mathbf{S}^\text{A}}; {\mathbf{S}^\text{B}};{\mathbf{S}^\text{C}}]$. The target output ${{\widehat{y}}_{3}}(t)$ is set as the magnitude of input frequencies as a function of time (orange dashed line in \textbf{Figure \ref{fig:input}}a). 

\medskip
Next, in a separate training, we train the same origami reservoir to identify different input amplitudes -- using the input magnitude of level 2 as pattern I, level 1 as pattern II, and level 4 as pattern III -- and collect the corresponding reservoir state matrix ${S} =[{\mathbf{S}^\text{I}}; {\mathbf{S}^\text{II}}; {\mathbf{S}^\text{III}}]$). Note that the input frequency is constant at 4Hz. The target output is set up similarly (black dashed line in Figure \ref{fig:input}a). 
By completing these two sets of training, we obtain two separate sets of readout weights ${{W}_{amp}}$ and ${{W}_{freq}}$.

\medskip
Figures \ref{fig:input}(a) summarize the training results of frequency and amplitude pattern identification, respectively. One can see that the origami reservoir can be successfully trained to output the correct frequencies and magnitudes. For comparison, we calculate another set of outputs without the origami reservoir by applying the readout weights only to the bottom four origami vertices' displacement (whose time responses are closest to the shake input). The corresponding results are shown by blue lines, which exhibit no input identification capability. This comparison proves the necessity of an origami reservoir with its nonlinear dynamics to accomplish computing and machine learning. 

\begin{figure}[b!]
    \centering
    \includegraphics[scale=0.95]{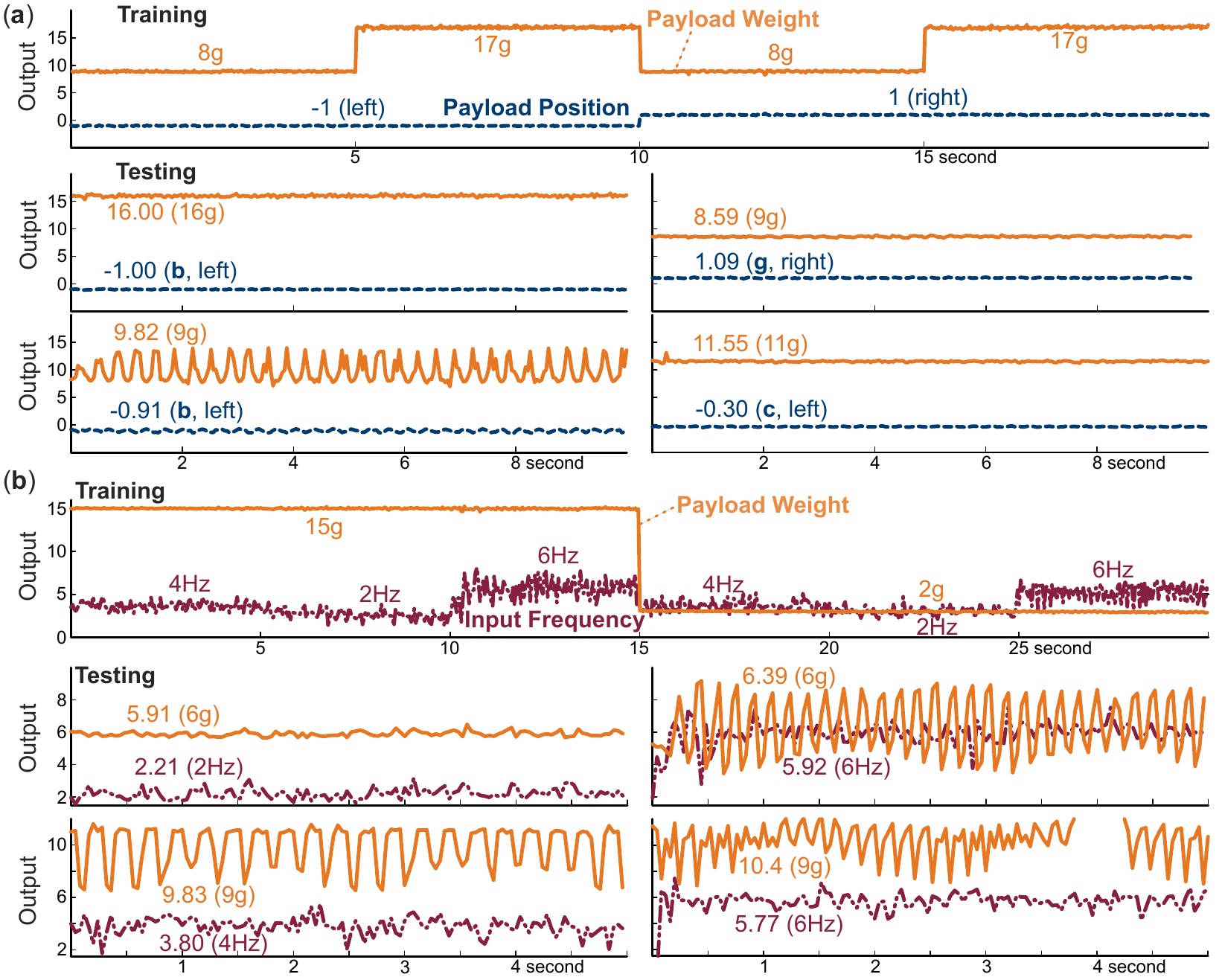}
    \caption{Multi-tasking with origami reservoir. (a, top) Training results for simultaneous payload weight and payload position estimation. In this case, we place an 8-gram training payload and then a 17-gram one on position ``a'' of the origami reservoir, each for 5 seconds in sequence. Then, they are put on position ``h.'' (a, bottom) The two concurrent outputs of origami reservoir under four testing conditions, when $<$magnitude, position$>$ is chosen as $<$16g, pb$>$, $<$9g, pg$>$, $<$9g, pb$>$, $<$11g, p3$>$. (b, top) Training results for simultaneous payload weight and input frequency recognition. In this case, we place the payload of 15g on position ``a'' and excited the origami using 4, 2, and 6Hz input frequencies, each for 5 seconds in sequence. Then we repeat the process with the 3-gram payload. (b, bottom) Two groups of reservoir outputs under four prediction cases when $<$magnitude, frequency$>$ is chosen as $<$6g, 2Hz$>$, $<$6g, 6Hz$>$, $<$9g, 4Hz$>$, and $<$9g, 6Hz$>$.}
    \label{fig:multi-task}
\end{figure}

\medskip
Once the training is complete, we test if the origami reservoir can recognize an unknown input signal with a random combination of the three frequencies and magnitudes. For example, Figure \ref{fig:input}(c) shows two input patterns produced with a random sequence of three frequencies and amplitudes, each lasting for a random duration. Once we apply the two sets of pre-trained readout weights ${{W}_{amp}}$ and ${{W}_{freq}}$ to the origami's vertice's displacements from these random excitations (a few examples shown in Figure \ref{fig:input}d), we can obtain the reservoir's prediction on the corresponding input parameters (Figure \ref{fig:input}e, f). 
To eliminate the influences of high-frequency fluctuation in the data, we sample and average the reservoir output every 0.2 s as the final result (solid red line in Figure \ref{fig:input}e,f), which is close to the true input frequencies and magnitudes.
%
%The frequency is fixed to 4Hz in amplitude pattern identification while we adjust the input number of amplitude slightly to make the real vibration amplitude remains roughly the same in frequency pattern identification, as shown in Figure \ref{fig:input signal pattern}d. 

\begin{figure}[b!]
    \centering
    \includegraphics[scale=0.95]{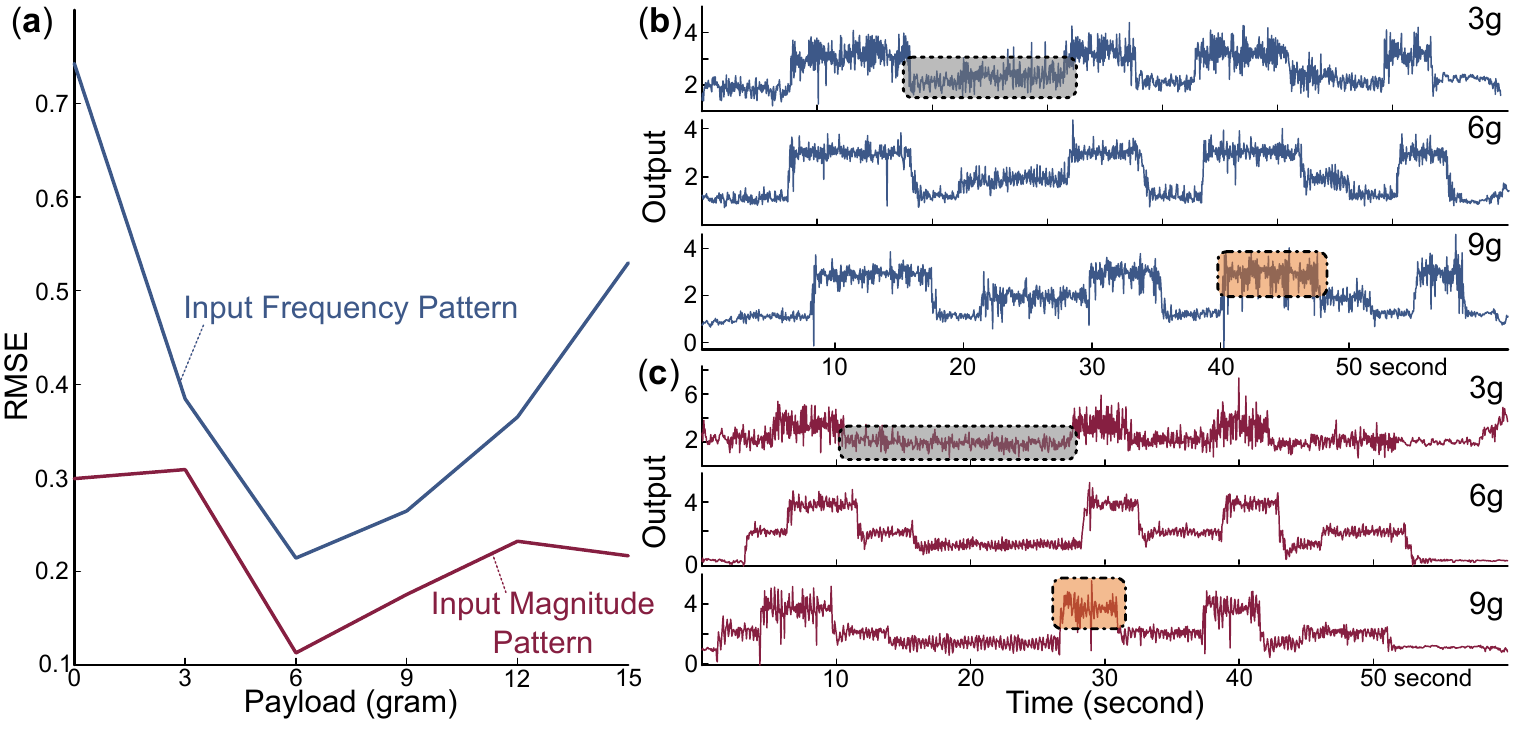}
    \caption{Influence of payload weight on the accuracy of input patterns recognition. (a) Relationship between RMSE of pattern recognition and the payload weight. (b, c) Reservoir outputs for input frequency and magnitude recognition with the 3g, 6g, and 9g payload, respectively. The true input parameters are the same as shown in Figure \ref{fig:input}(b).}
    \label{fig:weight_effect}
\end{figure}

\subsection{Multi-Tasking}
Multi-tasking in physical reservoirs has been achieved with emulation  (i.e., emulating several nonlinear filters with the same state vectors) \cite{bhovad2021physical}, but it has yet to be proved with more complex intelligent behaviors. Here, we use two case studies to explore multi-tasking in the origami reservoir and examine whether it can extra two types of information from the same state vectors. In the first case study, we task the origami reservoir to estimate the payload weight and position simultaneously. In the second case, we train the origami to recognize payload weight and input frequency simultaneously.

\medskip
Such multi-tasking requires more state vectors for training. For the first case study, four groups of state vectors are assembled for training: including the payload of 8 or 17 grams located at positions ``a'' or ``h.'' Therefore, the reservoir state vector matrix for training is $\mathbf{S} =[{\mathbf{S}^{8g, pa}}; {\mathbf{S}^{17g, pa}}; {\mathbf{S}^{8g, ph}}; {\mathbf{S}^{17g, ph}}]$, each component consisting of 5 seconds of data. Two target outputs are defined according to the real value of the payload weight and position (\textbf{Figure \ref{fig:multi-task}}), and each target output will give a set of readout weights.

\medskip
Training for payload weight and position simultaneously generates near-perfect fits to the targeted outputs (Figure \ref{fig:multi-task}a). Then we apply the two sets of readout weights to other reservoir state vectors from random payload weight and position combinations, such as  $<$16g, position b$>$, $<$9g, position b$>$, $<$9g, position g$>$, $<$11g, position c$>$). The results show that the origami reservoir can correctly classify the payload position and, at the same time, accurately estimate its weight with less than 10\% errors. Estimating an unknown payload's weight is more straightforward than classifying its position. One can add more position data to training. However, adding more training data under this setting of position target function do not promise to increase the accuracy rate of classification.

\medskip
For the second case study of concurrent payload weight and input frequency recognition, training requires six groups of state vectors, including 3 and 15-gram payloads, as well as 4, 2, and 6 Hz input frequencies (Figure \ref{fig:multi-task}c). Therefore, $\mathbf{S} =[{\mathbf{S}^{15g, 4Hz}}; {\mathbf{S}^{15g, 2Hz}}; {\mathbf{S}^{15g, 6Hz}}; {\mathbf{S}^{3g, 4Hz}}; {\mathbf{S}^{3g, 2Hz}}; {\mathbf{S}^{3g, 6Hz}}]$), each components containing 5 seconds of data. The two target outputs are set up similarly as shown in Figure \ref{fig:multi-task}(b).

\medskip
The training results for input frequency recognition show some fluctuations, but those for payload weight estimation are consistent. Similarly, we obtain two sets of readout weights from the training and then apply them to state vectors from four randomly selected input settings: $<$6g, 2Hz$>$, $<$6g, 6Hz$>$, $<$9g, 4Hz$>$, and $<$9g, 6Hz$>$. Surprisingly, even though the training result does not clearly separate the 2Hz and 4Hz input frequencies, the origami reservoir still precisely predicts the testing frequency under a different payload.

\medskip
In multi-perception task, the influence of different observing elements on the performance of each perception task should be noted. In general, we observe that the payload weight strongly influences the multi-tasking performance. For example, for the input frequency and magnitude recognition task, the origami reservoir's prediction is more accurate when the payload is near the same weight as the origami itself (\textbf{Figure \ref{fig:weight_effect}}a). Taking a closer look at the reservoir's output when the payload is 3, 6, and 9 grams, one can find it difficult to distinguish the reservoir's output between lower input frequencies (2 and 4Hz) and lower input amplitude (level 1 and 2 in shaker controller, as shown in data circled with the dashed lines in the first plot of Figure \ref{fig:weight_effect}b,c). On the other hand, if the payload becomes heavier at 9 grams, more chaotic data occur (data circled with dashed-dotted lines in the third plot).
%
%We have discussed the influence of frequency on payload prediction previously. As payload is a significant part for future intelligent robot, it is also important to explore the influence of payload on accuracy of frequency/amplitude identification. The result shows that identification is more accurate when payload is the same weight as the origami itself (Figure \ref{fig:payload influence}a). 

\begin{figure}[t!]
    \centering
    \includegraphics[]{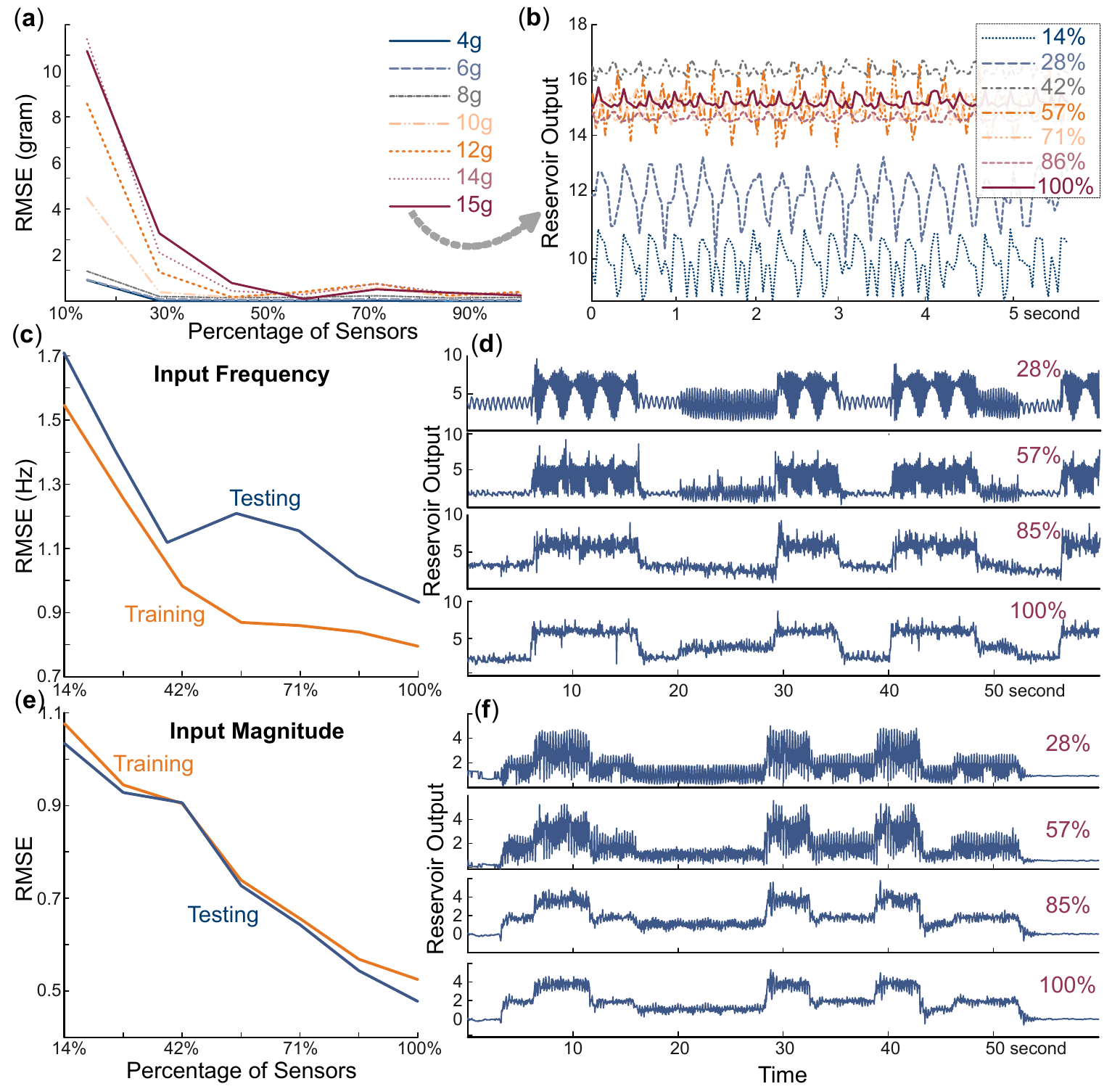}
    \caption{Influence of reservoir dimensionality (aka., number of vertices displacement) on information perception tasks performance. (a) The RSME of the weight estimation task for 4, 6, 8, 10, 12, 14, and 15-gram payloads using different amounts of vertices displacements. The payload is located in position ``a,'' with a 4Hz input frequency. (b) The reservoir output corresponding to the 15-gram payload and different computing dimensions. (c, e) The RSME for input frequency and amplitude pattern recognition based on various percentages of vertices displacements, respectively. (d, f) The reservoir outputs for the predicted frequency and amplitude patterns, respectively, when 8, 16, 24, and 28 nodal displacements are used. The true input parameters are the same as shown in Figure \ref{fig:input}(b).}
    \label{fig:sensor}
\end{figure}

\subsection{Reduced Reservoir Dimensionality}
In the case studies above, we collected all 28 vertices displacements to construct the state vector matrices ($\mathbf{S}$).
Such design can, however, lead to a complicated mechatronic setup if one wants to use embedded sensors to measure these reservoir states.
Therefore, it is vital to uncover the influence of reducing reservoir dimension on information perception performance by reducing the number of vertices displacement used in these tasks. We analyze the rooted mean square error of payload weight estimation and frequency/amplitude recognition tasks based on randomly selected 4 (14\% of total vertices), 8 (28\%), 12 (42\%), 16 (57\%), 20 (71\%), 24 (86\%), and 28 (100\%) vertices displacements.

\medskip
\textbf{Figure \ref{fig:sensor}}(a) shows that the error of payload weight estimation rapidly decreases when we increase the dimension of computing data. About 30\% of vertices displacement is sufficient to obtain an accurate prediction. This observation is consistent with the authors' previous simulation results \cite{wang2022experimental} {[ref]}. 
Figure \ref{fig:sensor}(b) details some examples of the reservoir output with a 15-gram payload. The output from only 14\% of computing nodes deviates significantly from the actual payload weight. But when we use 28\% of the vertices (or nodal) displacement, the averaged reservoir output quickly shifts to the true value, although there is a significant fluctuation in the output data. As we use more vertices displacements, such fluctuation decreases.

\medskip
The situation is different for the input parameter recognition tasks. The errors in input frequency and amplitude recognition decrease more gradually with increasing computing dimension. As a result, roughly 50\% of the vertex displacement is necessary for accurate predictions (Figure \ref{fig:sensor}c and \ref{fig:sensor}e, respectively). When closely comparing the reservoir outputs with 14\%, 57\%, 71\%, and 100\% of the vertices displacements, one can see that the origami reservoir cannot filter undesired signals and recognize the input parameters precisely with less than 60\% of nodal displacements. Therefore, input pattern recognition demands more computing power than payload weight estimation.

\section{Summary and Discussion}
Via extensive experimentation, we demonstrate that origami structures harbor sufficient physical reservoir computing capacity to perform intelligent information perception tasks like input recognition, payload weight estimation, payload position classification, and a combination of two tasks. In all these tasks, the computation or machine learning occurs in the physical vibrations of the origami, and only a simple linear regression is required for training. We also obtained insights into how to set up the origami reservoir for better intelligence performance. Here we discuss some important observations and conclusions. 

\medskip
First, we need two payloads --- one relatively light compared to the origami and the other heavy --- to train the readout weights for payload weight estimation. The accuracy of payload weight estimation is high when the weight of other payloads falls between the two training values, and the input frequency is low (1 – 3.5 Hz).
Second, precisely estimating the payload position is challenging. Therefore, we use the origami reservoir as a classifier to predict whether the payload is on the left or right half. This classification task is more likely to succeed when the payload is much heavier than the origami, and the input frequency is relatively high (3 – 6Hz). Under these conditions, origami can exhibit significantly non-uniform deformation to separate the response from different payload positions.
Therefore, a moderate input frequency (3Hz) is the ideal choice to estimate the payload magnitude and position simultaneously (aka. multi-tasking). 
Finally, the origami reservoir accomplished input pattern recognition tasks, a pre-requisite embedded controlling and behavior switch. 

\medskip
Besides these three information perception tasks, we prove that multi-tasking is achievable by applying two target functions to one group of state vector matrices (e.g., estimating payload weight and position; or predicting payload weight and input frequency). However, the increased training burden will naturally sacrifice accuracy, especially when some tasks negatively impact the performance of others (e.g., in Figure \ref{fig:weight_effect}). Thus, it is crucial to consider the trade-off between multi-tasking and desired accuracy. 

\medskip
Despite these promising results, several challenges arise during the experiment effort. The most significant is the robustness and repeatability. The origami exhibits plastic deformation during repeated vibration, so the resting positions of vertices inevitably drift slightly over time. The origami and the payload could receive minor disturbance by accident, which will also influence the results of nonlinear projection. Therefore, ensuring an identical experimental setup between training and testing is vital. 
The other challenging phenomenon is the fluctuations in reservoir outputs, especially in the input recognition tasks. One possible explanation is that we use the vertices’ displacement as state vectors, which shows less nonlinearity than other state variables, such as creases angle. In conclusion, a successful setup of an origami reservoir for information perception depends on robust training of readout weights, consistent design of the physical platform, careful choice of reservoir state vectors, and a sufficiently large dimension for computing. Overall, the results of this study are stepping stones for building more sophisticated and practical intelligent tasks for constructing advanced materials and robots with mechano-intelligence. 

% Acknowledgements
\medskip
\textbf{Acknowledgements} \par %delete if not applicable))
The authors acknowledge the support from the National Science Foundation (CMMI-1933124).
% References
\medskip

% Use the following code if you wish to generate your bibliography with BibTeX;
% replace the string "MSP-template" below with the name(s) of
% the BibTeX data base(s) you want to use.
% The resulting bibliography-output (the content of the .bbl file)
% must be pasted back into this file before submission.
% Please also include your BibTeX data base file(s) in your submission
% so that we can re-run BibTeX if necessary.
%
\bibliographystyle{MSP}
\bibliography{reference}

\end{document}